\documentclass[%
reprint,
 amsmath,amssymb,
 aps,
pra,
]{revtex4-1}

\usepackage{color}
\usepackage[table]{xcolor}
\usepackage{graphicx}
\usepackage{dcolumn}
\usepackage{bm}
\usepackage{natbib}
\usepackage{comment}
\usepackage[normalem]{ulem} 

\usepackage{multirow}

\newcommand{\req}[1]{Eq.\,(\ref{eq:#1})}

\begin{document}


\title{Radiation reaction friction: Resistive material medium
}

\author{Martin Formanek}
\email{martinformanek@email.arizona.edu}
\author{Andrew Steinmetz}
\author{Johann Rafelski}
\affiliation{Department of Physics, The University of Arizona,
Tucson, AZ 85721-0081, USA
}

\date{September 9, 2020}

\begin{abstract}
We explore a novel method of describing the radiation friction of particles traveling through a mechanically resistive medium. We introduce a particle motion induced matter warping along the path in a manner assuring that charged particle dynamics occurs subject to radiative energy loss described by the Larmor formula. We compare our description with the Landau-Lifshitz-like model for the radiation friction and show that the established model exhibits non-physical behavior. Our approach predicts in the presence of large mechanical friction an upper limit on radiative energy loss being equal to the energy loss due to the mechanical medium resistance.  We demonstrate that mechanical friction due to strong interactions, for example of quarks in quark-gluon plasma, can induce significant soft photon radiation.

\end{abstract}

\maketitle

\section{Introduction} 

We study the motion of particles subject to a covariant mechanical friction force (MFF) caused by the presence of a material medium. In general, in the presence of any force, a charged particle emits radiation, a result obtained by Larmor considering properties of Maxwell\rq s equations. Emitted radiation complements the MFF as an induced radiation friction force (RFF). MFF will be used as an insightful model to learn how to accommodate the dynamics of radiation reaction force, {\it i.e.\/} radiation reaction (RR). 

To the best of our knowledge, all prior covariant studies of RR employed the Lorentz force (LF) due to externally prescribed electromagnetic (EM) fields. Considered in the context of a Lorentz-Maxwell system of dynamical equations, it is well known that RR is an unsolved problem. The advantage of our approach is that we can focus on a better understanding of the effect of the radiation reaction on the mechanically accelerated particle, without need to reconcile the LF with Maxwell field dynamics. 

We choose a MFF force which reduces to the familiar form of Newton\rq s friction force in the non-relativistic limit in the case of linear relativistic motion. In Sect.~\ref{sec:covfriction} we find that the relativistic generalization of Newtonian friction has a unique form used also in the study of Brownian motion~\cite{Dunkel:2005}. For constant MFF we evaluate stopping distance, rapidity shift, and stopping power, which provide background for the later study of motion including RFF. 

In this work we introduce, as a mathematical tool for making RR consistent with special relativity, a matter warping model along the particle path by particle acceleration and parameterized by the proper time of the particle. Matter warping is not a field as it is known only along the path of an accelerated particle. When borrowing tools of differential geometry we therefore prefer to speak of a warped  matter metric rather than a curved space-time metric. There have been earlier efforts to modify the space-time metric for accelerated particles spearheaded by Caianiello~\cite{Caianiello:1989wm}, whose work was driven by the postulate of a maximal proper acceleration. For a recent review see~\cite{Torrome:2018zck}. Our objective is  clearly very different  even though some of the methods are similar as we explore the physical environment of a resistive medium  and in particular its warping due to accelerated motion.

To summarize the  theoretical advantages of using the material medium: 
\begin{enumerate}
\item Unlike in the space-time empty of matter (vacuum) case, the presence of a material medium provides a reference frame against which we measure particle motion. Hence the covariant form of MFF depends on the particle 4-velocity as well as the 4-velocity of the medium.
\item When a particle experiences energy loss due to RFF, this occurs in the model always at the expense of the well-defined relative motion with respect to the medium.
\item The specific form of the LF does not enter and thus the inconsistency between the EM field equations and the description of charged particle motion  subject to EM-force, see discussion on p. 745 in Jackson~\cite{Jackson:1999}, is not introduced. We tacitly employ the Maxwell field equations when characterizing the magnitude of radiative energy loss for an accelerated charged particle as it is well known from Larmor\rq s work.
\item A metric warped by particle acceleration within a material medium can have the simple interpretation as being due to local material response to particle motion.
\end{enumerate}

This study is constrained to exploring RR for accelerated charged particle motion within matter only. Our exploration is limited to warping along the particle worldline due to acceleration. Although we use similar mathematical methods as in the case of curved space-time we don't actually claim that the space-time is curved.  We expect that path warping method can help advance the generalization of  our approach to the case of accelerated charged particle motion in vacuum, which motivates the introduction of this method in this work. However, understanding of space-time geometry warping outside the particle worldline maybe  also required. Such a new theoretical framework is beyond  our current scope, has not yet ben formulated and is not needed here to advance the understanding of RR we develop.

The path-warped  method description of RR avoids the introduction of higher order derivatives into the equations of motion, see discussion on p. 393 in Panofsky-Phillips~\cite{Panofsky:1962}. The Lorentz-Abraham-Dirac (LAD) equation\rq s higher order derivative term was introduced  to assure orthogonality of the equation of motion with respect to 4-velocity. This term leads to causality challenges and runaway solutions. There are different interpretations of this term: In some derivations of RR this term is introduced ad-hoc as necessary to assure constancy of the speed of light $u^2 =$ const~\cite{Rohrlich:1965cl,Barut:1980aj}; There is an effort to derive it  based on the Lorentz-force of the regularized self-field, see for example the work in Ref.~\cite{Bild:2019dlu} which is further developing Dirac\rq s derivation of LAD~\cite{Dirac:1938nz}, we return to this issue below in Section~\ref{ssec:LADfric}. This controversial term does not appear in our formulation.

In view of these difficulties, Landau-Lifshitz~\cite{LL:1962} proposed an iterative scheme using dynamical equations to eliminate higher derivatives. We show in Sect.~\ref{sec:LL} that for a particle decelerated in the medium the Landau-Lifshitz-like model of radiation friction predicts non-physical behavior for particle motion. This alone demonstrates the need to find another method to incorporate RR into in-medium particle dynamics. The validity for both LAD and LL description is restricted to the classical domain of particle behavior \cite{DiPiazza:2008}.

In Sect.~\ref{sec:modmetr} we show that the \lq pure\rq\ Larmor-RR term proportional to particle 4-velocity can be a natural consequence of a suitable matter warping along the particle path. To achieve this we  characterize matter warping by an explicit dependence of the metric on the particle path and its acceleration. This naturally satisfies the requirement discussed by Langevin~\cite{Langevin:1911} that \lq\lq being accelerated\rq\rq\ marks the body in a distinct way in that the magnitude of time dilation depends on the history of acceleration. We note that Langevin\rq s remarks do not depend on the particle moving only in vacuum, they retain in full their meaning for motion in resistive material medium as well. We will return to the more difficult case of motion in vacuum in the follow-up work.

The present reformulation of RR contributes as well to a better understanding of LAD, which has been interpreted as the interaction of the charged particle with its own radiation field~\cite{Bild:2019dlu}. However, both classical and quantum particles do not move within their own Coulomb fields. With this in mind, we posit that such particles should not be allowed to move under the influence of their own radiation fields as well. It is a textbook exercise, see sect.\;29.4 in Ref\;\cite{Rafelski:2017}, to show that a charged accelerated particle in its instantaneous co-moving frame generates both the Coulomb field and the radiative field and there is no relative motion with an observer required to establish this. This is because, unlike velocity, acceleration has an absolute meaning, and only in the instantaneous co-moving frame does the acceleration 4-vector have the pure space-like format $a^\mu=(0,\vec a)$. Use of matter warping naturally prevents the particle from  being accelerated by its own radiation field.

In Sect.~\ref{sec:solution} we implement the numerical solution for the equations of motion and compare it to the motion without radiation friction. We show that in the limit of high rapidity and/or mechanical friction strength the radiation friction loss is at most matching the mechanical friction energy loss. We briefly consider application of such a model for high energy particle collisions. 

\section{Friction force in medium}\label{sec:covfriction}
In this section we describe motion of a particle under the influence of a covariant friction force in a resistive medium. The form of the force is such that it reduces to the Newtonian friction in the non-relativistic limit. We derive the expressions for stopping distance, rapidity shift and loss of energy and momentum. This force is present for both neutral and charged particles, and the energy loss manifests itself in general as heat dissipation.
\subsection{Covariant equation of motion}
The covariant equation of motion and the friction force are given by
\begin{equation}\label{eq:motion}
\dot{u}^\mu = \frac{1}{m}\mathcal{F}^\mu, \quad \mathcal{F}^\mu = \frac{r}{c}P^\mu_{\ \ \nu}\eta^\nu\,,
\end{equation}
where $r$ is the strength of the friction and $P^\mu_{\ \ \nu}$ is the projector on the orthogonal direction to 4-velocity $u^\mu$ of the particle
\begin{equation}
P^\mu_{\ \ \nu} = \delta^\mu_\nu - \frac{u^\mu u_\nu}{c^2}\,.
\end{equation}
This ensures that our friction force is automatically orthogonal to the 4-velocity. Finally, we denote the 4-velocity of the medium as $\eta^\nu$. For general choice of the 4-velocities
\begin{equation}
\eta^\mu = (\gamma_\text{M}c,\gamma_\text{M} \pmb{v}_\text{M}), \quad u^\mu = (\gamma c, \gamma \pmb{v}), 
\end{equation}
\begin{equation}
\eta \cdot u = \gamma_\text{M} \gamma c^2 (1 - \pmb{\beta}_\text{M} \cdot \pmb{\beta}\;,
\end{equation}
the zeroth and spatial components of the equation of the motion \req{motion} read
\begin{align}
\label{eq:friction0th}\gamma \frac{d\gamma}{dt} &= \frac{r}{mc}\gamma_\text{M}(1 - \gamma^2(1-\pmb{\beta}_\text{M}\cdot \pmb{\beta}))\;,\\
\gamma \frac{d\gamma\pmb{\beta}}{dt} &= \frac{r}{mc}\gamma_\text{M}(\pmb{\beta}_\text{M} - \gamma^2(1-\pmb{\beta}_\text{M} \cdot \pmb{\beta})\pmb{\beta})\,.
\end{align}
The energy balance, given by \req{friction0th}, is overall negative when
\begin{equation}
\pmb{\beta}_\text{M} \cdot \pmb{\beta} < \beta^2\,,
\end{equation}
which means the particle loses energy due to friction. In the opposite case the medium is moving faster than the particle and the particle is being accelerated to match the velocity of the medium. When $\pmb{\beta}_\text{M} = \pmb{\beta}$ the particle reaches an equilibrium state of rest with respect to the medium and the friction force completely disappears. 

Let\rq s explore further the behavior of the friction force in the rest frame of the medium when $\pmb{\beta}_\text{M} = 0$, $\gamma_\text{M} = 1$ or in 4-vector notation
\begin{equation}
\eta^\mu = (c, 0, 0,0), \quad \eta \cdot u = \gamma c^2\;.
\end{equation}
In this case the components of \req{motion} are
\begin{align}
\label{eq:zerothrfm}\gamma \frac{d\gamma}{dt} &= \frac{r}{mc} (1-\gamma^2)\,,\\
\label{eq:spatialrfm}\frac{d}{dt}(\gamma \pmb{\beta}) &= - \frac{r}{mc} \gamma \pmb{\beta}\,. 
\end{align}
We can always orient our coordinate system so that the initial velocity of the particle coincides with one of the coordinate axes. Consider that the particle enters the medium in the $x$-direction. The perpendicular velocity then remains zero for the duration of the particle\rq s travel and the motion is entirely one-dimensional. In terms of rapidity $y$ satisfying
\begin{equation}\label{eq:rapidity}
\gamma = \cosh y, \quad \gamma \beta = \sinh y, \quad \beta = \tanh y\,,
\end{equation}
we can re-write both equations \req{zerothrfm} and \req{spatialrfm} as
\begin{equation}\label{eq:dydtnoRR}
\frac{dy(t)}{dt} = - \frac{r}{mc}\tanh y(t)\,,
\end{equation}
which has a solution for initial rapidity $y(0) = y_0$ 
\begin{equation}\label{eq:ytnoRR}
y(t) = \text{Arcsinh}\left(\sinh(y_0)\exp\left(-\frac{r}{mc}t \right)\right)\,.
\end{equation}
Note that if the velocity of the particle with respect to the medium is small, $\beta \ll 1$, the equation of motion \req{spatialrfm} becomes
\begin{equation}
m \frac{dv}{dt} = - \frac{r}{c} v \,,
\end{equation}
which is a familiar Newtonian friction force linearly proportional to velocity. In order to account for friction with more complicated behavior than linear dependence on velocity we need to replace the constant $r$ with a function of relative velocity $r(\eta\cdot u)$, which is manifestly a Lorentz scalar. In such case the solution \req{ytnoRR} would have to be replaced by a numerical solution of \req{dydtnoRR} with a specific function $r(y)$. The friction strength $r$ is also in general a function of the medium density. 

\subsection{Distance traveled and rapidity shift}
Rewriting the solution $y(t)$ in \req{ytnoRR} as
\begin{equation}
\sinh y(t) = \sinh y_0 \exp\left(-\frac{r}{mc}t\right)\,,
\end{equation}
and using $\sinh y = \gamma \beta$, we can find solution for $\beta(t)$ as
\begin{equation}
\beta(t) = \frac{\gamma_0\beta_0}{\sqrt{\exp\left(\frac{2r}{mc}t\right) + \gamma_0^2\beta_0^2}}\,.
\end{equation} 
The distance traveled in the rest frame of the medium is given by the integral
\begin{align}
x(t) =& x_0 + \int_0^t \beta(t')cdt' \nonumber\\[10pt]
=& x_0 + \int_0^t \frac{\gamma_0\beta_0}{\sqrt{\exp\left(\frac{2r}{mc}t'\right) + \gamma_0^2\beta_0^2}}cdt'\,,
\end{align}
which can be evaluated as 
\begin{align}
x(t) = &x_0 + \frac{mc^2}{r}\left(\text{Arctanh}(\beta_0) - \text{Arctanh}\left(\beta(t)\right)\right)\nonumber\\[5pt]
\label{eq:xtnoRR}
=& x_0 + \frac{mc^2}{r}(y_0 - y(t))\;.
\end{align}
The total distance traveled $D$ until the particle comes to a stop $y(t_s) = 0$ at time $t_s$ is simply
\begin{equation}\label{eq:stopdist}
D = \left. (x(t) - x_0)\right|_{y(t_s) = 0} = \frac{mc^2}{r}y_0\,.
\end{equation}
By inverting \req{xtnoRR} we can obtain $y$ as a function of $x$
\begin{equation}\label{eq:yx}
y(x) = y_0 - \frac{r}{mc^2}(x - x_0)\,.
\end{equation}
The rapidity shift per change in distance $dy/dx$ is therefore a constant
\begin{equation}\label{eq:stoppowy}
\frac{dy}{dx} = - \frac{r}{mc^2}\,.
\end{equation}
This is the mechanical rapidity shift caused by the medium\rq s resistance. In the case of charged particle motion we also need to account for the additional radiation rapidity shift effect. The description of this contribution is the main focus of Sect.~\ref{sec:LL} and Sect.~\ref{sec:modmetr}.

\subsection{Energy and momentum loss}
We now can consider the stopping power in terms of the change of energy $E = mc^2\gamma$ per unit of distance
\begin{equation}\label{eq:dydx}
\frac{dy}{dx} = \frac{d\text{Arccosh}(\gamma)}{dx} = \frac{1}{\sqrt{\gamma^2-1}}\frac{d\gamma}{dx} = \frac{1}{mc^2\sinh y}\frac{dE}{dx}\,.
\end{equation}
Therefore by substituting \req{dydx} into \req{stoppowy}, we obtain
\begin{equation}\label{eq:dEdx}
\frac{dE}{dx} = - r \sinh y(x)\,.
\end{equation}
Clearly, when rapidity reaches zero in the rest frame of the medium, the particle energy stops changing as expected. Another way to write this expression uses
\begin{equation}
\sinh y = \gamma \beta = \frac{p}{mc}\,,
\end{equation}
where $p = m\gamma v$ is particle\rq s momentum. Then
\begin{equation}
\frac{dE}{dx} = - \frac{r}{mc} p\,,
\end{equation}
and conversely using the relativistic expression for energy $E = \sqrt{m^2c^4 + p^2c^2}$
\begin{equation}
\frac{dp}{dx} = - \frac{r}{mc^3}E\,.
\end{equation}
Finally, energy and momentum can be expressed as 
\begin{align}
E = mc^2 \gamma &= mc^2 \cosh y\,,\\
p = mc\gamma\beta &= mc \sinh y\,, 
\end{align}
which we can evaluate either as a function of laboratory time or position, using the solutions for rapidity $y(t)$ \req{ytnoRR} and $y(x)$ \req{yx}, respectively.

\section{Radiation friction}\label{sec:LL}
This section introduces: a) the \lq standard model\rq\ of RFF, the Lorentz-Abraham-Dirac (LAD) equations of motion for description of RR; and b) Landau-Lifshitz reduction of LAD differential order. We apply this procedure to our problem of the charged particle moving in a resistive medium. We present Landau-Lifshitz-like (LLL) equations of motion for both Newtonian friction and friction with strength generally dependent on $\eta\cdot u$ and discuss their behavior. We show that the LLL model leads to non-physical behavior for the particle motion.
\subsection{LAD radiation friction in vacuum}\label{ssec:LADfric}
The unresolved question of the consistent description of accelerated charged particle motion including its radiation and radiation friction is now well over a century old. Indeed, the power radiated by such particle was first described by Larmor at the end of the 19th century~\cite{Larmor:1897rad}. In a covariant form
\begin{equation}\label{eq:larmor}
P = m \tau_0 \dot{u}^2\,,
\end{equation}
where the characteristic time $\tau_0$ reads
\begin{equation}
\tau_0 = \frac{2}{3} \frac{e^2}{4\pi\varepsilon_0 \varepsilon_r mc^3} = \frac{2}{3} \frac{\alpha\hbar}{mc^2\varepsilon_r}
\approx 6.26 \times 10^{-24}\text{ s}\,,
\end{equation}
for an electron in vacuum, where $\alpha\approx1/137.036$, $\varepsilon_r = 1$. If we add a corresponding momentum change to \req{motion} the equation of motion reads
\begin{equation}\label{eq:larmorfirst}
\dot{u}^\mu \stackrel{?}{=} \frac{1}{m}\mathcal{F}^\mu + \tau_0 \dot{u}^2 \frac{u^\mu}{c^2}.
\end{equation}
We see that this expression does not preserve $u^2 = c^2$ because $\dot{u} \cdot u \neq 0$. 
Further work by Abraham~\cite{Abraham:1903cl}, Dirac~\cite{Dirac:1938nz}, and Lorentz~\cite{Lorentz:1952th} resulted in the formulation of the LAD equation for accelerated charged particle motion 
\begin{equation}\label{eq:LAD}
\dot{u}^\mu = \frac{1}{m}\mathcal{F}^\mu + \tau_0 \left(\ddot{u}^\mu + \dot{u}^2 \frac{u^\mu}{c^2}\right)\,,
\end{equation}
where the term proportional to the second derivative of 4-velocity is the so-called Schott term. 

We consider this term controversial as briefly outlined in the introduction. In some very well known LAD derivations it is added ad-hoc to ensure $u^2 =$ const~\cite{Rohrlich:1965cl,Barut:1980aj}. However, search to justify this terms presence is offered in derivations based on the Lorentz-force of the regularized self-field~\cite{Bild:2019dlu}. This derivation clarifies that to justify the Schott term one must posit that a particle can experience its own self-force. This cannot be the case if classical dynamics arises in a limiting process of quantum physics, where such a self-force for matter particles (fermions) is not possible (see \cite{Merzbacher:1970}, p.533).

The Shott term is the origin of the theoretically unwelcome second order proper-time derivative of the 4-velocity. This leads to a number of unresolved issues with initial conditions, causality, run-away and pre-accelerated solutions~\cite{Spohn:2004, Poisson:1999}. This field of study remains very active till this day, with recent publications exploring particles of finite extent and their point limit~\cite{Gralla:2009} and improving as noted the methods explored in~\cite{Bild:2019dlu}. Lastly, and of fundamental importance, the Schott term is the critical obstacle in many failed efforts to find a variational principle formulation of RR  -- in absence of such a formulation the  charged particle dynamics with RR lacks conservation laws required in a complete and consistent description.

Currently there are two main approaches aiming to resolve the issue of the LAD formulation. The first is to impose appropriate boundary and asymptotic conditions on the solution so that the non-physical solutions are discarded~\cite{Plass:1961}. This approach is difficult to implement for problems which require numerical solutions. Instead we will compare our results with a second approach of the Landau-Lifshitz (LL) model~\cite{LL:1962} which approximates the LAD equation by iterating the acceleration due to external force and expanding into powers of the  parameter $\tau_0$. This approach reduces the order of the equation of motion and thus resolves the known issues of the LAD formulation at the expense of approximation of the the full radiation reaction.

Spohn~\cite{Spohn:2000} showed that LAD restricted onto a physical manifold produces the LL series, so both approaches lead in certain environment to the same dynamics. However, this is not the case in the study of electron stopping by a frontal light plane wave, see Ref.\cite{Hadad:2010mt}. This occurs because a traveling light wave front creates a quasi-material edge for an incoming particle. This resembles, but is not exactly the same as, the case of the material medium we look at next.
 
\subsection{Landau-Lifshitz-like RR in medium}
\subsubsection{Constant material friction}
For our system in the zeroth order in $\tau_0$, the acceleration is given by the external force
\begin{equation}
\dot{u}^\mu_{(0)} = \frac{r}{mc} P^\mu_{\ \ \nu}\eta^\nu\,.
\end{equation}
By substituting this expression to the radiation friction term in \req{LAD} we obtain for the second derivative of 4-velocity
\begin{multline}
\ddot{u}^\mu_{(0)} = \frac{r}{mc}\dot{P}^\mu_{\ \ \nu}\eta^\nu = \frac{r}{mc}\left(-\frac{\dot{u}^\mu_{(0)} (\eta \cdot u)}{c^2} - \frac{u^\mu (\dot{u}_{(0)} \cdot \eta)}{c^2}\right)\\
= - \frac{r^2}{m^2c^4}\left(P^\mu_{\ \ \nu}\eta^\nu (\eta \cdot u) + u^\mu(\eta \cdot P \cdot \eta)\right)\,,
\end{multline}
and for the square of acceleration in the zeroth order in $\tau_0$
\begin{equation}
\dot{u}^2_{(0)} = \frac{r^2}{m^2c^2}(\eta \cdot P \cdot P \cdot \eta) = \frac{r^2}{m^2c^2}(\eta \cdot P \cdot \eta)\,,
\end{equation}
because of the property of the projector $P^2 = P$. We see that the Larmor term cancels with one of the two terms arising from the Schott term and the final Landau-Lifsthitz-like (LLL) equation of motion reads
\begin{equation}
\dot{u}_{(1)}^\mu = \frac{r}{mc} P^\mu_{\ \ \nu}\eta^\nu - \tau_0 \frac{r^2}{m^2c^4} (\eta \cdot u) P^\mu_{\ \ \nu}\eta^\nu\,.
\end{equation}
The zeroth component of this equation in the rest frame of the medium is
\begin{equation}
\gamma \frac{d\gamma}{dt} = \frac{r}{mc}(1-\gamma^2)-\tau_0 \frac{r^2}{m^2c^2}\gamma(1-\gamma^2)\,,
\end{equation}
and in terms of rapidity \req{rapidity} we have
\begin{equation}\label{eq:dydtLL}
\frac{dy}{dt} = - \frac{r}{mc}\tanh y + \tau_0 \frac{r^2}{m^2c^2}\sinh y\,.
\end{equation}
In the second term of \req{dydtLL} we see a reversal in the sign. The effect of radiation friction is then, up to the first power in $\tau_0$, to increase the energy of the particle experiencing deceleration in a medium. Moreover, the rapidity has to satisfy 
\begin{equation}\label{eq:LLcondition}
y < \text{arccosh}\left(\frac{mc}{\tau_0 r} \right)\,,
\end{equation}
otherwise the radiation friction overpowers the mechanical friction. Teitelboim et al. \cite{Teitelboim:1980} argue that both, LAD equation and its LLL reduction  are unjustified  when the radiation friction force is comparable to the driving external force. Violation of this condition in the medium  results in a runaway solution which is clearly an unacceptable behavior. As the Lorentz force is not relevant in the material friction case, the incompatibility must originate more fundamentally with the LAD extension. \\

\subsubsection{Variable material friction in LLL approach}
The derivation above assumes that the radiation friction $r$ is constant. If we evaluate LLL model for non-constant $r$ as a function of $\eta\cdot u$, then the covariant equation of motion up to first order in $\tau_0$ is
\begin{multline}
\dot{u}^\mu_{(1)} = \frac{r}{mc} P^\mu_{\ \ \nu}\eta^\nu +\\
+ \tau_0 \frac{r}{m^2c^2}\left(-r \frac{\eta \cdot u}{c^2} + \frac{dr}{d(\eta \cdot u)}(\eta \cdot P \cdot \eta)\right) P^\mu_{\ \ \nu}\eta^\nu\,,
\end{multline}
and the zeroth component in the rest frame of the medium is
\begin{equation}
\gamma \frac{d\gamma}{dt} = \frac{r}{mc}(1-\gamma^2) + \tau_0 \frac{r}{m^2c^2}\left(-r\gamma + \frac{dr}{d\gamma}(1-\gamma^2)\right)(1-\gamma^2)\,.
\end{equation}
Such LLL friction term has a chance of having negative contribution to energy if 
\begin{equation}\label{eq:inequality}
\frac{dr}{d\gamma} < - \frac{r\gamma}{\gamma^2 -1}\,.
\end{equation}
This cannot happen in the non-relativistic limit, because $dr/d\gamma$ would have to go to minus infinity. The terms exactly cancel when 
\begin{equation}\label{eq:zeroRR}
r \propto \frac{1}{\sqrt{\gamma^2 -1}} = \frac{1}{\gamma\beta} = \frac{mc}{p}\,.
\end{equation}
In such a case there is no radiation friction according to the LLL approach. We introduce this example of mechanical friction with the friction coefficient $r$ depending inversely on momentum to present the singular case when radiation friction force disappears completely. For more realistic models of mechanical friction, when the coefficient $r$ grows with momentum, the inequality in \req{inequality} shows that the LLL terms add energy to the system. We conclude that the conventional LAD radiation reaction combined with LLL reduction of the order of differentiation produces an unacceptable description of radiation friction in a material medium, even when the friction strength is an arbitrary function of relative velocity. 

\section{Matter warping}\label{sec:modmetr}

Here we propose an alternative radiation friction model  for the case of motion in matter medium. We show that formally we can introduce radiation friction in medium through matter warping while keeping the form of the covariant Larmor formula. This allows us to formulate the dynamics without higher order derivatives and with self-consistent formula for the magnitude of acceleration. As the driving force we take the covariant mechanical friction force. We establish equations of motion for such a system in the warped matter model and evaluate stopping power.

\subsection{General considerations}
As already noted in the introduction, our warped path formulation is not requiring exploration of space-time beyond the particle path: We start with the equation of motion with only the Larmor term present
\begin{equation}\label{eq:curvedlarmor}
\dot{u}^\mu = \frac{1}{m}\mathcal{F}^\mu + \tau_0 \dot{u}^2 \frac{u^\mu}{c^2}\;.
\end{equation}
This specific form of the equation of motion is our choice of collective medium response model to accelerated charged particle motion, guided by the visual similarity with \req{larmorfirst}, mathematical tractability, and interpretability of the results {\it i.e.} mathematical simplicity and beauty. Instead of adding a second order derivative Schott term we assume path-warped metric allowing us to impose the condition
\begin{equation}\label{eq:u2=c2}
u^2 \equiv  g_{\mu\nu}u^\mu u^\nu = c^2
\end{equation}
i.e. the 4-force remains orthogonal to 4-velocity. To see this we multiply \req{curvedlarmor} by $g_{\mu\nu}u^\nu$ to obtain
\begin{equation}\label{eq:larmordotuu}
\dot{u} \cdot u = \tau_0 \dot{u}^2\,.
\end{equation}
If we use this identity in \req{curvedlarmor} we derive an expression which is explicitly orthogonal to $u_\mu$ 
\begin{equation}\label{eq:motion_modmetr}
\dot{u}^\mu - (\dot{u} \cdot u) \frac{u^\mu}{c^2} = \frac{1}{m}\mathcal{F}^\mu\,,
\end{equation}
suggesting that the covariant friction is proportional to the product of 4-velocity and 4-acceleration $\dot{u} \cdot u$, which is normally zero. Upon differentiating the condition $u^2 = c^2$ \req{u2=c2} with respect to proper time we obtain
\begin{equation}\label{eq:dotuu}
\dot{u} \cdot u = - \frac{1}{2} \frac{dg_{\mu\nu}}{d\tau} u^\mu u^\nu\,,
\end{equation}
which is a condition for components of the metric. \req{curvedlarmor}--\req{dotuu} provide  a consistent characterization of the magnitude of acceleration $\dot{u}^2$. The square of this expression reads
\begin{equation}
\dot{u}^2 = \frac{1}{m^2}\mathcal{F}^2 + \tau_0^2 \frac{\dot{u}^4}{c^2}\;, 
\end{equation}
which is a quadratic equation for $\dot{u}^2$ with solutions
\begin{equation}
\dot{u}^2 = \frac{c^2}{2\tau_0^2}\left(1 \pm \sqrt{1-4\frac{\tau_0^2}{c^2}\frac{\mathcal{F}^2}{m^2}}\right)\,.
\end{equation}
We take the minus sign as the physical solution, because it reduces in the limit $\tau_0 \rightarrow 0$ to the usual expression $\dot{u}^2 = \mathcal{F}^2/m^2$. This expression can be further simplified to
\begin{equation}\label{eq:dotu2}
\dot{u}^2 = \frac{2\mathcal{F}^2/m^2}{1+\sqrt{1-4\frac{\tau_0^2}{c^2}\frac{\mathcal{F}^2}{m^2}}}\,.
\end{equation}
It is worth noting that as $\mathcal{F}^2 \rightarrow 0$ then also $\dot{u}^2 \rightarrow 0$ and conversely as $\mathcal{F}^2 \rightarrow -\infty$ the growth of $\dot{u}^2$ is damped.

\subsection{Specific warped matter model}
In order to avoid mixing the spatial and time components we will assume that the metric is diagonal and in our 1D situation we choose a parametrization
\begin{equation}
g_{\mu\nu} = \text{diag}(f_0^2,-f^2,-1,-1)\,.
\end{equation}
In the following we will suppress the two trivial degrees of freedom. The proper time of the particle is by definition
\begin{equation}
d\tau = \frac{1}{c}\sqrt{g_{\mu\nu}\frac{dx^\mu}{dt}\frac{dx^\nu}{dt}} dt= \sqrt{f_0^2 - \frac{f^2}{c^2}\left(\frac{dx}{dt}\right)^2}dt\,,
\end{equation}
where we took the position 4-vector as $x^\mu = (ct,x)$. We can then perform a coordinate transformation from $dt$ and $dx$ to measurable quantities
\begin{equation}\label{eq:transformation}
dt_\text{lab} = f_0 dt, \quad dx_\text{lab} = f dx\;,
\end{equation}
which simplifies the increment of proper time to
\begin{equation}
d\tau = \sqrt{1-\frac{1}{c^2}\left(\frac{dx_\text{lab}}{dt_\text{lab}}\right)^2}dt_\text{lab}\,.
\end{equation}
If we define the true physical velocity of the particle as
\begin{equation}
v \equiv \frac{dx_\text{lab}}{dt_\text{lab}} = \frac{f}{f_0}\frac{dx}{dt}\,,
\end{equation}
we can write the gamma factor in the usual form
\begin{equation}
\gamma = \frac{dt_\text{lab}}{d\tau} = \frac{1}{\sqrt{1-v^2/c^2}}\,.
\end{equation}
Note that the transformation \req{transformation} is a transformation to flat space coordinates, as can be seen by evaluating
\begin{equation}
ds^2 = g_{\mu\nu}dx^\mu dx^\nu = c^2dt_\text{lab}^2 - dx_\text{lab}^2\,.
\end{equation}

The coordinate 4-velocity $u^\mu$ that enters our equation of motion is given by 
\begin{equation}\label{eq:umu}
u^\mu \equiv \frac{dx^\mu}{d\tau} = \gamma \left(\frac{dct}{dt_\text{lab}}, \frac{dx}{dt_\text{lab}} \right) = \gamma \left(\frac{c}{f_0}, \frac{v}{f}\right)\;.
\end{equation}
Here $u^\mu$ is the quantity in terms of which the equation of motion is formulated, thus it is tempting to look at it as the actual 4-velocity. However, a more robust theoretical framework is needed to give $u^\mu$ a physical meaning and this will be required for the solution of the vacuum case. This expression satisfies $u^2 = c^2$ \req{u2=c2} as expected and energy of the particle is given by the usual expression
\begin{equation}
E = \gamma m c^2\,.
\end{equation}
For the 4-velocity of the medium we can write analogically
\begin{equation}
\eta^\mu = \gamma_\text{M} \left(\frac{c}{f_0},\frac{v_\text{M}}{f}\right)\,,
\end{equation}
preserving $\eta^2 = c^2$. Therefore the RHS of the equation of motion \req{motion_modmetr} in the rest frame of the medium reads
\begin{align}
\frac{1}{m}\mathcal{F}^\mu = \frac{r}{mc}P^\mu_{\ \ \nu}\eta^\nu =& \frac{r}{mc}\left(\eta^\mu - \frac{u^\mu}{c^2}(\eta \cdot u)\right) \nonumber\\
\label{eq:force}
=& \frac{r}{mc}\left(\frac{c}{f_0}(1-\gamma^2), - \frac{\gamma^2}{f}v\right)\,,
\end{align}
which is equivalent to rescaling the zeroth component of the force by $f_0$ and spatial component by $f$. Note that $r$ can be in general a function of $\eta \cdot u$ to account for more complicated mechanical friction than Newtonian friction, but this fact does not modify our derivation and holds true throughout the current section (Sec.~\ref{sec:modmetr}). Finally, we can evaluate the dot product $\dot{u}\cdot u$ using \req{dotuu}
\begin{equation}\label{eq:dotuueval}
\dot{u}\cdot u = -\frac{1}{2}\frac{df_0^2}{d\tau}\frac{\gamma^2c^2}{f_0^2} + \frac{1}{2}\frac{df^2}{d\tau}\frac{\gamma^2v^2}{f^2} \equiv -A \gamma^2 c^2 + B \gamma^2 v^2\,,
\end{equation} 
where we denoted
\begin{equation}
A \equiv \frac{d}{d\tau}\ln f_0, \quad B \equiv \frac{d}{d\tau} \ln f\,.
\end{equation}
Now we are prepared to establish the equations of motion.
\subsection{Radiation energy loss in our model}
If we substitute the 4-velocity \req{umu}, the force \req{force}, and the dot product \req{dotuueval} to the equation of motion \req{motion_modmetr} we obtain for the zeroth component
\begin{equation}\label{eq:zeroth1}
\frac{d}{d\tau}\left(\frac{\gamma}{f_0}\right) + (A\gamma^2c^2-B\gamma^2v^2)\frac{\gamma}{f_0 c^2} = \frac{r}{mcf_0}(1-\gamma^2)\,.
\end{equation}
The first term can be further expanded
\begin{equation}
\frac{d}{d\tau}\left(\frac{\gamma}{f_0}\right) = \frac{\gamma}{f_0}\frac{d\gamma}{dt_\text{lab}} - \frac{1}{f_0^2}\gamma \frac{df_0}{d\tau} = \frac{\gamma}{f_0}\frac{d\gamma}{dt_\text{lab}} - \frac{\gamma}{f_0} A\,.
\end{equation}
Finally, by substituting back to \req{zeroth1} and multiplying by $f_0/\gamma$ we have an expression for the change in gamma factor 
\begin{equation}\label{eq:zeroth2}
\frac{d\gamma}{dt_\text{lab}} = \frac{r}{mc\gamma}(1-\gamma^2) + A(1-\gamma^2) + B\gamma^2\beta^2\,.
\end{equation}
Similarly for the spatial component of \req{motion_modmetr}
\begin{equation}\label{eq:spatial1}
\frac{d}{d\tau}\left(\frac{\gamma v}{f}\right) + (A\gamma^2c^2-B\gamma^2v^2)\frac{\gamma v}{f c^2} = - \frac{r\gamma^2}{mf}v\,.
\end{equation}
For the first term in \req{spatial1} we evaluate the derivative
\begin{align}
\frac{d}{d\tau}\left(\frac{\gamma v}{f}\right) =& \gamma \frac{d\gamma}{dt_\text{lab}} \frac{v}{f} - \frac{1}{f^2} \frac{df}{d\tau} \gamma v + \frac{\gamma^2}{f}\frac{dv}{dt_\text{lab}}\nonumber\\
=& \gamma \frac{d\gamma}{dt_\text{lab}}\frac{v}{f} - \frac{\gamma}{f}B v + \frac{\gamma^2}{f}\frac{dv}{dt_\text{lab}}\,.
\end{align}
If we use \req{zeroth2} for change in $\gamma$ and substitute back to \req{spatial1} then after several cancellations we obtain
\begin{equation}
\frac{\gamma^2}{f}\frac{dv}{dt_\text{lab}} = - \frac{r}{mc}\frac{v}{f} - (A-B)\frac{\gamma v}{f}\;,
\end{equation}
which if we multiply by $f/\gamma^2 c$ becomes
\begin{equation}
\frac{d\beta}{dt_\text{lab}} = - \frac{r}{mc}\frac{\beta}{\gamma^2} - (A-B)\frac{\beta}{\gamma}\,.
\end{equation}
We can use identity $1 + \gamma^2\beta^2 = \gamma^2$ to further simplify the zeroth part \req{zeroth2} and write the components of the covariant equation of motion in a final form
\begin{align}
\label{eq:zerothfinal}\frac{d\gamma}{dt_\text{lab}} &= \frac{r}{mc\gamma}(1-\gamma^2) + (A-B)(1-\gamma^2)\,,\\
\label{eq:spatialfinal}\frac{d\beta}{dt_\text{lab}} &= - \frac{r}{mc}\frac{\beta}{\gamma^2} - (A-B)\frac{\beta}{\gamma}\,,
\end{align}
which are mutually equivalent. Although the metric $g_{\mu\nu}$ is specified by two unknowns, $A$ and $B$, the dynamics of the particle motion depends only on their difference 
\begin{equation}
A - B = \frac{d}{d\tau}\ln \frac{f_0}{f}\,,
\end{equation}
which means that any arbitrary factor re-scaling the whole metric does not change the motion. Therefore we can set either $A$ or $B$ to zero and evaluate the other without any loss of generality.
\subsection{Stopping power and limiting cases}
We assume that the more general equation of motion \req{motion_modmetr} is in the form of \req{curvedlarmor} where the friction term is given by the Larmor formula. Combining expressions for $\dot{u} \cdot u$ in \req{larmordotuu} and \req{dotuu} we can equate
\begin{equation}
\tau_0 \dot{u}^2 = - \frac{1}{2} \frac{dg_{\mu\nu}}{d\tau} u^\mu u^\nu\,,
\end{equation}
where the self-consistent magnitude of acceleration $\dot{u}^2$ was evaluated in \req{dotu2} and the right hand side is given in our metric as \req{dotuueval}
\begin{equation}
\frac{2\tau_0\mathcal{F}^2/m^2}{1+\sqrt{1-4\frac{\tau_0^2}{c^2}\frac{\mathcal{F}^2}{m^2}}} = -A \gamma^2 c^2 + B \gamma^2 v^2\,.
\end{equation}
The square of the external force can be computed using equation \req{force}
\begin{equation}
\frac{1}{m^2}\mathcal{F}^2 =\frac{1}{m^2}g_{\mu\nu}\mathcal{F}^\mu\mathcal{F}^\nu = \frac{r^2}{m^2}(1-\gamma^2)\,.
\end{equation}
Notice that this expression does not depend on the metric and is equal to the square of the external force in the flat space-time. As discussed above we can set $A = 0$ and evaluate $B$
\begin{equation}
B = \frac{-2\tau_0 \frac{r^2}{m^2c^2}}{1 + \sqrt{1+4\frac{\tau_0^2}{c^2}\frac{r^2}{m^2}\gamma^2\beta^2}}\,,
\end{equation}
where we used the identity $1-\gamma^2 = -\gamma^2\beta^2$. Using this solution in the equations of motion \req{zerothfinal} and \req{spatialfinal} yields
\begin{align}
\label{eq:zerothspecific}
\frac{d\gamma}{dt_\text{lab}} &= \frac{r}{mc\gamma}(1-\gamma^2) + \frac{2\tau_0 \frac{r^2}{m^2c^2}(1-\gamma^2)}{1 + \sqrt{1+4\frac{\tau_0^2}{c^2}\frac{r^2}{m^2}\gamma^2\beta^2}}\;,\\
\label{eq:spatialspecific}
\frac{d\beta}{dt_\text{lab}} &= - \frac{r}{mc}\frac{\beta}{\gamma^2} - \frac{2\tau_0 \frac{r^2}{m^2c^2}}{1 + \sqrt{1+4\frac{\tau_0^2}{c^2}\frac{r^2}{m^2}\gamma^2\beta^2}} \frac{\beta}{\gamma}\,.
\end{align}
From \req{zerothspecific} we can calculate the radiation energy loss in powers of $\tau_0$ 
\begin{equation}
\left. \frac{dE}{dt_\text{lab}}\right|_\text{RF} = mc^2 \left. \frac{d\gamma}{dt_\text{lab}}\right|_\text{RF} = \tau_0 \frac{r^2}{m}(1-\gamma^2) + O(\tau_0^3)\;,
\end{equation}
which matches the covariant Larmor energy loss formula \req{larmor} with 4-acceleration given purely by the external force.

In terms of stopping power $dE/dx$ we can convert \req{zerothspecific} to 
\begin{equation}
\frac{dE}{dx_\text{lab}} = \left. \frac{dE}{dx_\text{lab}}\right|_\text{M} - \frac{\frac{2\tau_0}{mc} \left(\left.\frac{dE}{dx_\text{lab}}\right|_\text{M}\right)^2\coth y}{1 + \sqrt{1 + 4 \frac{\tau_0^2}{m^2c^2}\left(\left.\frac{dE}{dx_\text{lab}}\right|_\text{M}\right)^2}}\,,
\end{equation} 
where the $dE/dx_\text{lab}|_\text{M}$ is the stopping power caused by the medium friction given by \req{dEdx}. The two important limiting cases are determined by critical mechanical stopping power
\begin{equation}\label{eq:criticalstoppow}
\left. \frac{dE}{dx_\text{lab}}\right|_\text{crit} \equiv \frac{mc^2}{c\tau_0} = \frac{ 3}{ 2}\; \frac{(mc^2)^2}{\alpha \hbar c} \approx 0.27 \varepsilon_r \text{ MeV/fm}\,,
\end{equation}
where the value given is for an electron in an environment with relative permittivity $\varepsilon_r$. 

If mechanical stopping power is much higher than the critical stopping power the radiation friction part of the stopping power is approximately
\begin{equation}\label{eq:supercritical}
\left. \frac{dE}{dx_\text{lab}}\right|_\text{RF} \approx \left. \frac{dE}{dx_\text{lab}}\right|_\text{M} \coth y\,.
\end{equation}
In the opposite case, when mechanical stopping power is much less than the critical stopping power,
\begin{equation}
\left. \frac{dE}{dx_\text{lab}}\right|_\text{RF} \approx \left. \frac{dE}{dx_\text{lab}}\right|_\text{M} \left(\frac{\left.\frac{dE}{dx_\text{lab}}\right|_\text{M}}{\left.\frac{dE}{dx_\text{lab}}\right|_\text{crit}}\right) \coth y\,.\
\end{equation}
Note that if $y \rightarrow 0$ then the stopping power in medium also goes to zero as $\sinh y$, so the expression is well behaved.
\section{Motion examples}\label{sec:solution}
With the dynamics developed in the previous section (Sect.~\ref{sec:modmetr}) we can evaluate the motion of the radiating charged particles and compare to the LLL model (Sect.~\ref{sec:LL}) and to the motion without any radiation friction (Sect.~\ref{sec:covfriction}). This section presents numerical solutions for the particle motion in each of the three situations with the underlying Newtonian mechanical friction force. We show that our model unlike the LLL model increases, as expected, the energy loss due to radiation friction. This additional energy loss can at most match the mechanical friction loss in the medium in the limit of high rapidity or friction strength. Finally, we discuss possible experimental applications of our model for high energy particle collisions.
\subsection{Solution of dynamical equations}
\begin{figure}
	\centerline{\includegraphics[width=1.08\linewidth]{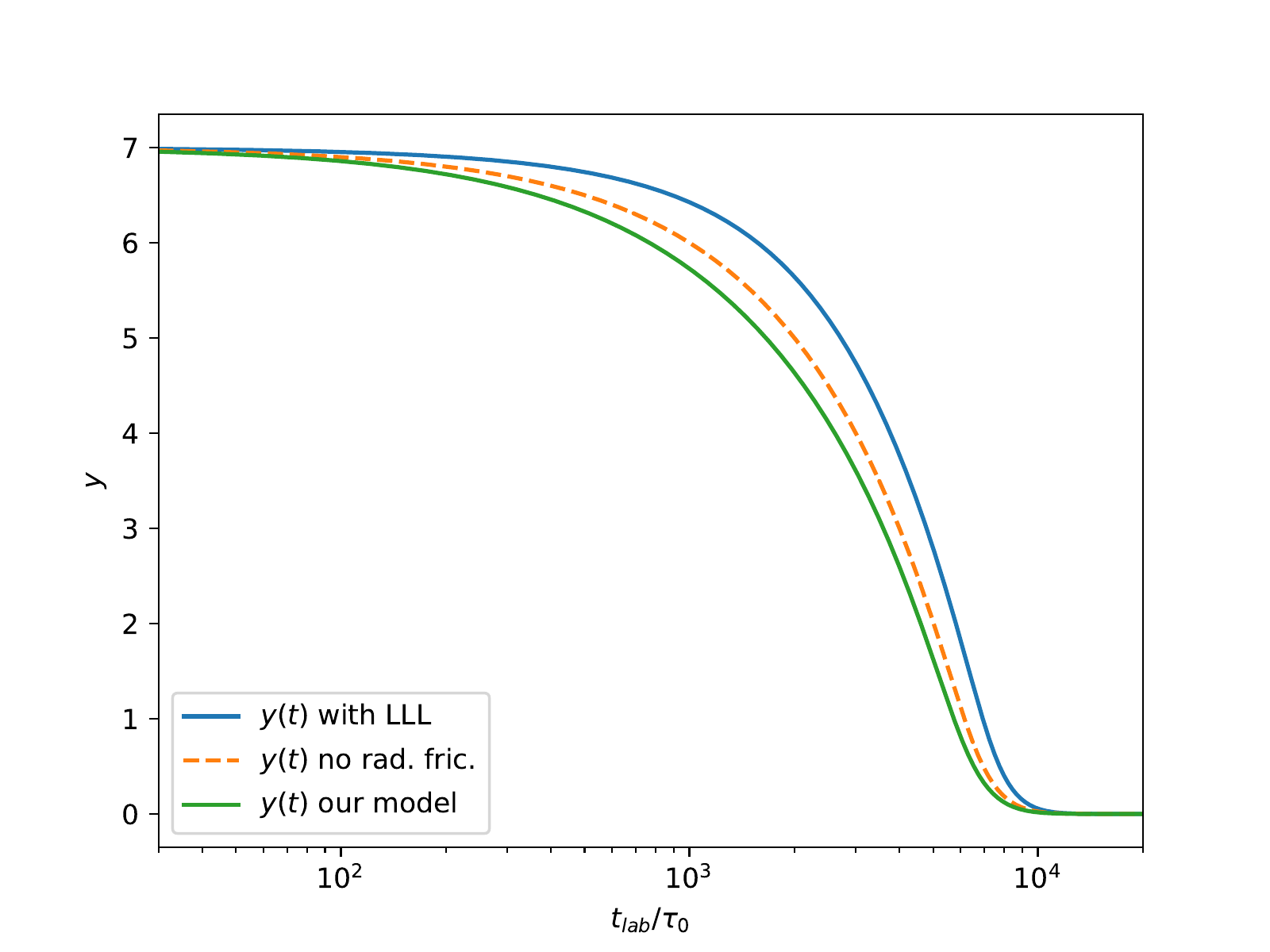}}
	\caption{\label{fig:y_t} Rapidity as a function of time for our model, LLL model and analytical solution without radiation friction. Initial condition $y(0) = 10^{-3}$ and $\tilde{r} =10^{-3}$.}
\end{figure}

\begin{figure}
	\centerline{\includegraphics[width=1.1\linewidth]{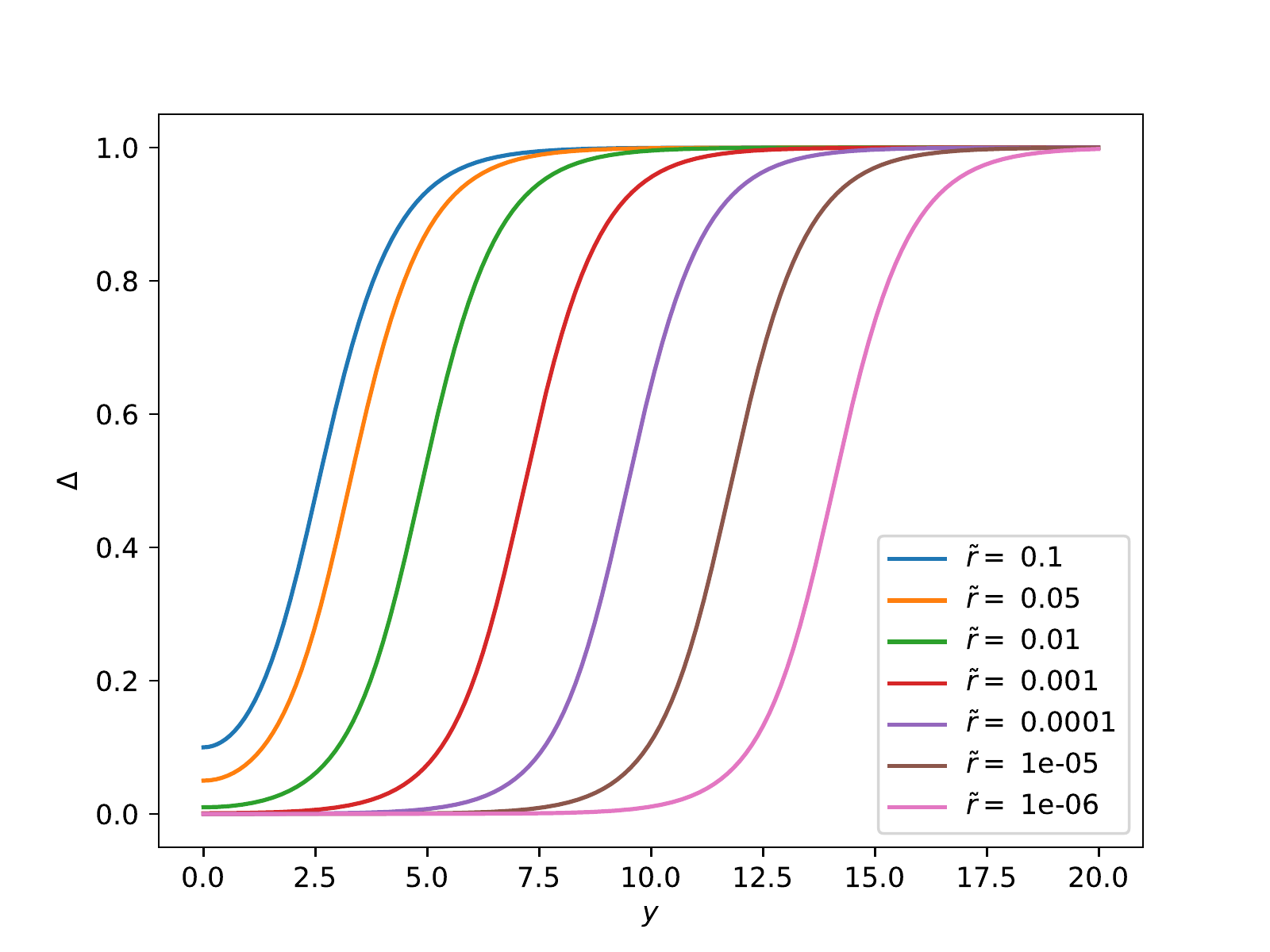}}
	\caption{\label{fig:delta} Relative change of the stopping power (see \req{dydxwithRR}) if we consider radiation friction for selected friction strengths $\tilde{r}$ and range of rapidities $y$.}
\end{figure}

Let us define a unitless friction strength
\begin{equation}
\tilde{r} \equiv \frac{r\tau_0}{mc}=\frac{r}{\left. {dE}/{dx_\text{lab}}\right|_\text{crit}} \,,
\end{equation}
then we can write the equation of motion \req{spatialspecific} as
\begin{equation}
\frac{d\beta}{dt_\text{lab}/\tau_0} = - \tilde{r} \frac{\beta}{\gamma^2} - \frac{2\tilde{r}^2}{1+\sqrt{1+4\gamma^2\beta^2\tilde{r}^2}}\frac{\beta}{\gamma}\,.
\end{equation}
Switching to rapidity \req{rapidity} we obtain
\begin{equation}\label{eq:dydt}
\frac{dy}{dt_\text{lab}/\tau_0} = - \tilde{r}\tanh y - \frac{2\tilde{r}^2}{1+\sqrt{1+4\tilde{r}^2\sinh^2 y}} \sinh y. 
\end{equation}
For comparison the LLL model \req{dydtLL} gives us an equation for rapidity
\begin{equation}
\frac{dy}{dt_\text{lab}/\tau_0} = - \tilde{r}\tanh y + 2\tilde{r}^2\sinh y. 
\end{equation}
These expressions are suitable for numerical analysis. Let us consider motion with initial rapidity $y_0 = 7$ and $\tilde{r} = 10^{-3}$ to demonstrate the character of the solution. Figure \ref{fig:y_t} shows rapidity as a function of time for both our model and the LLL model as propagated by the RK4 integration scheme. Note that the condition \req{LLcondition} is satisfied to prevent runaway solutions in the LLL model. The dashed line indicates the analytical solution without any radiation friction \req{ytnoRR}. We see that in the LLL model, the particle decelerates more slowly than without radiation friction and in our model faster. The behavior of our model should match our intuition as the \lq correct\rq\ physical behavior where both sources of friction impede the particle\rq s motion.

Stopping power can be evaluated by diving the whole expression \req{dydt} with $\beta = \tanh y$ because $dx_\text{lab} = \beta c dt_\text{lab}$
\begin{align}
\frac{dy}{dx_\text{lab}/c\tau_0} =& -\tilde{r}\left(1 + \frac{2\tilde{r}\cosh y}{1+\sqrt{1+4\tilde{r}^2\sinh^2 y}} \right)\nonumber \\[5pt]
\label{eq:dydxwithRR}
\equiv& \tilde{r}(1 + \Delta)\,,
\end{align}
where distance is measured in units of $c\tau_0$ and $\Delta$ is the relative change from the case without any radiation friction. Fig. \ref{fig:delta} shows possible values for $\Delta$ for selected unitless friction strengths $\tilde{r}$ and range of rapidities $y$. We see that the radiation friction at most doubles the stopping power $dy/dx_\text{lab}$. 

\begin{figure}
	\centerline{\includegraphics[width=1.05\linewidth]{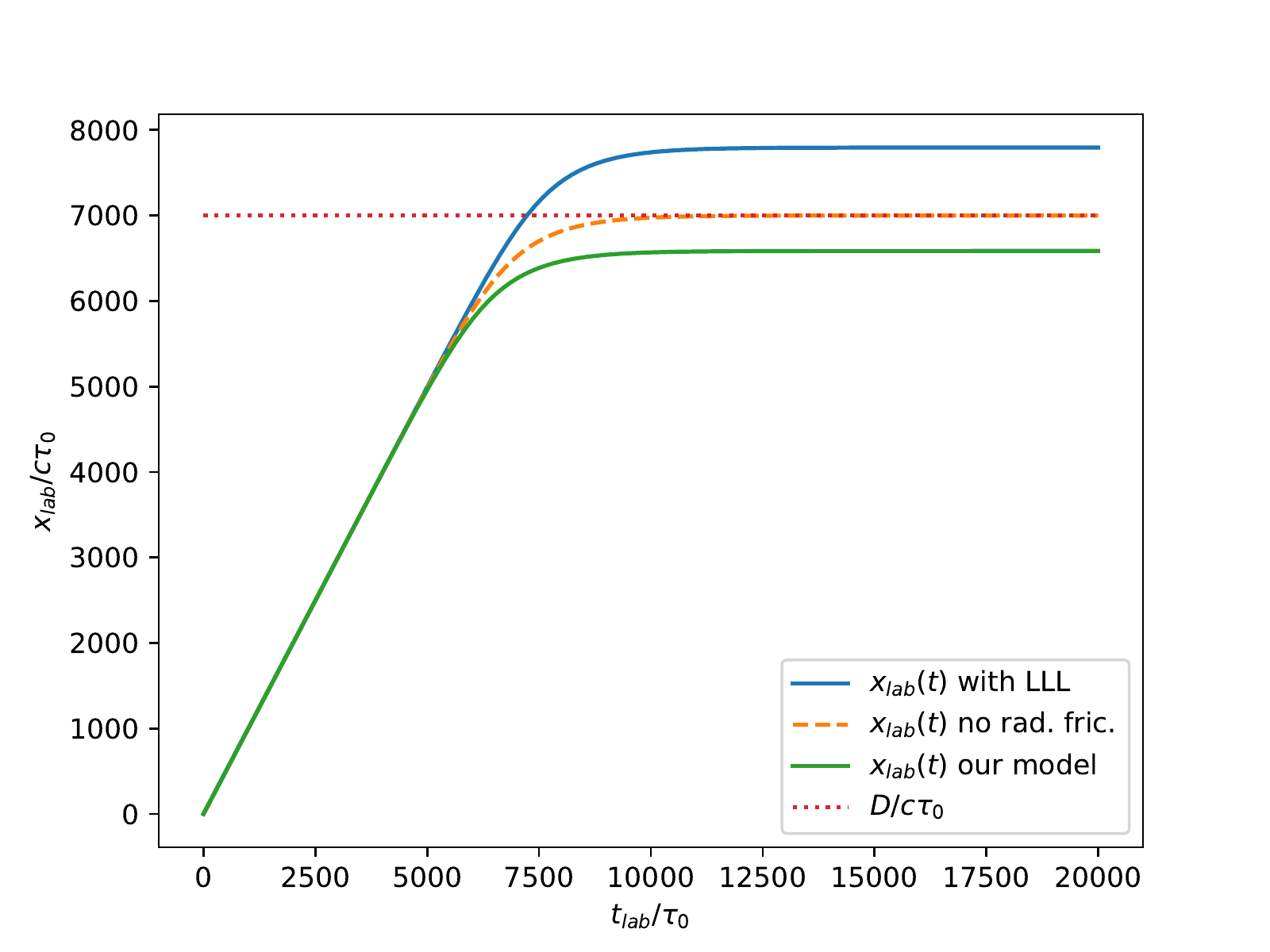}}
	\caption{\label{fig:x_t} Distance traveled by the particle as a function of time for initial rapidity $y_0 = 7$ and friction strength $\tilde{r} = 10^{-3}$. We show results with and without radiation friction and for the LLL model. Dotted line marks the stopping distance without radiation friction.}
\end{figure}

\begin{figure}
	\centerline{\includegraphics[width=1.1\linewidth]{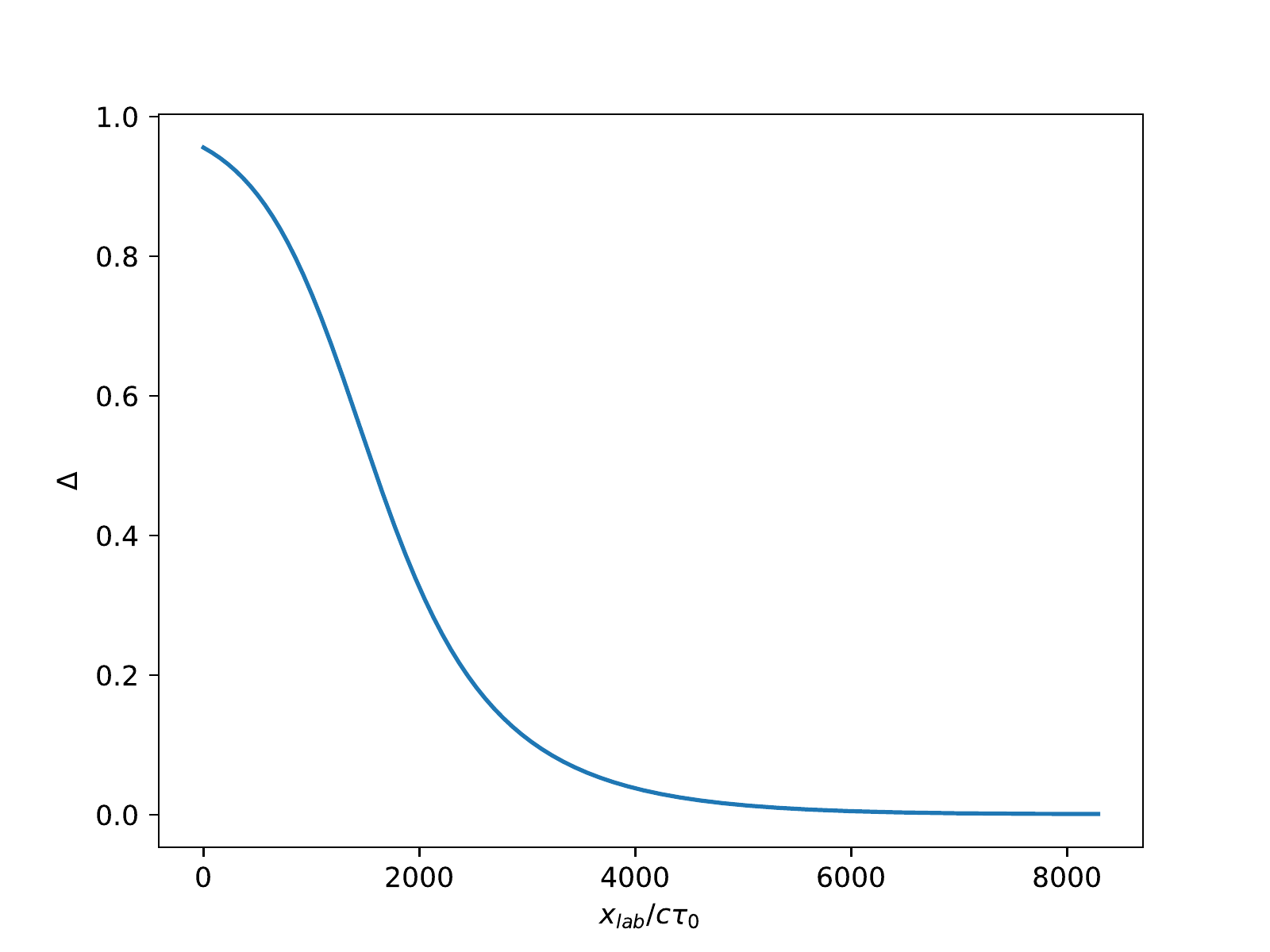}}
	\caption{\label{fig:deltaspec} Relative change of the stopping power (see equation \req{dydxwithRR}) as a function of distance for initial rapidity $y_0 = 10$ and friction strength $\tilde{r} = 10^{-3}$.}
\end{figure}

With our choice of parameters the stopping distance $D$ is in the case without radiation friction given by \req{stopdist}, which in unitless quantities reads
\begin{equation}
\frac{D}{c\tau_0} = \frac{y_0}{\tilde{r}} = 7 \times 10^3\,.
\end{equation}
As can be seen from the trajectories in Fig. \ref{fig:x_t} with radiation friction present, the particle in our model stops in a significantly shorter distance and for the LLL model it travels further. Figure~\ref{fig:deltaspec} demonstrates that initially the stopping power is doubled for high rapidities and as particle slows down the radiation friction contributes less and less. 

\subsection{Experimental verification}
From the expression for the critical stopping power \req{criticalstoppow} we see that a very high energy loss is needed to reach significant radiation friction. However, the typical stopping power of proton beams in a material medium is on the order of $1-100$ MeV/cm~\cite{Berger:1999} and this applies also to many other particles. This value is many orders of magnitude too small to induce sizable radiation energy production. This value is of course dependent on the density of the medium and varies with the energy of the particle. But even at the high end of the range, such stopping power is only a $10^{-11}$ fraction of the critical stopping power. 

We conclude that in normal materials with atomic structure, RR induced by the material stopping power is negligible, hence RR induced by MFF is negligible too. However, microscopically the particle motion is also experiencing Coulomb scattering off atomic nuclei of the medium with high accelerations, which at each scattering event contribute to the bremsstrahlung~\cite{Seltzer:1985}. In this work we do not consider these microscopic processes as we do not want to deal with EM forces presently.

Our RR effect is highly  relevant in the experimental environment of quark-gluon plasma, particularly in application to parton jet quenching processes. In the case of fast moving quarks the continuous medium covariant friction model is justified because quark-gluon plasma is successfully modeled as a fluid, beginning with the seminal work of Bjorken~\cite{Bjorken:1982qr}.
The value for the critical stopping power \req{criticalstoppow} for up- and down-quarks is
\begin{align}
\left. \frac{dE}{dx_\text{lab}}\right|_\text{crit} (u) &= 11.4 \varepsilon_r \text{ MeV/fm},\\
\left. \frac{dE}{dx_\text{lab}}\right|_\text{crit} (d) &= 204 \varepsilon_r \text{ MeV/fm},
\end{align}
where we used for the mass of the up-quark  $m_u = 2.2$ MeV/$c^2$, and of the down-quark $m_d = 4.7$ MeV/$c^2$, respectively. Since the up-quark has a higher fractional charge and a lower mass, the required critical mechanical stopping power is significantly lower when compared to the down-quark. Electrically charged quarks approaching critical mechanical friction would then emit significant amount of (soft) electromagnetic radiation. In addition there is  the possibility of superluminal Cherenkov radiation, and a further small contribution of acoustic wave production.  

According to work of Baier et.al~\cite{Baier:2000}  collisional stopping power of light quarks in a quark-gluon plasma at $T = 0.25$ GeV is on the order of $200$--$300$\,MeV/fm. Additional contributions arise from gluon-emission friction and other strong field effects. Certainly, the up-quark mechanical friction is supercritical leading to a significant EM radiation energy emission according to \req{supercritical}. A full study of  EM emissivity by quark jets would require incorporation of electrical permittivity of quark-gluon plasma and is well beyond our current scope. Excess soft photon emission in QGP has already attracted attention of the heavy-ion community. For a full discussion of relevant work see Section 3.3 in Ref.~\cite{Adamova:2019vkf}, most recent experimental results are found in Ref.\cite{Belogianni:2002}. Further effort to explore this effect in the context of our theoretical framework is warranted.

\section{Future work and conclusions} 
In the forthcoming work we hope to investigate other, less material environments. A study of EM interaction has, as we have mentioned, the challenge of reconciling the form of the Lorentz force with that of Maxwell\rq s equations. Therefore a training non-material problem in study of RR is an exploration of charged particle dynamics in presence of an external scalar force field. In this case the field dynamics providing the Larmor RR term is decoupled from the force form, which is non-material and thus reaching beyond the case considered in this work. We plan to return to this example after a short delay.

A further forthcoming training exercise is the study of RR for particles under influence of an externally prescribed constant electromagnetic field. There is in particular a very good reason to take a second look at the constant field case: The LAD or LL format of RR may not correctly describe the physical reality of a constant EM field. For example, a well known prediction of the LL model is that a particle linearly accelerated by an electric field (so-called hyperbolic motion) does not feel any radiation friction and yet produces radiation. This is due to the contribution of the Schott term in the equation of motion, which in this special case exactly balances out the Larmor term, a situation that is subject to ongoing discussion~\cite{Eriksen:2000,Eriksen:2002}. 

Even though a novel RR force patterned after this work requires establishment of consistency between the Maxwell field equations and the equation governing the particle motion we are optimistic that our proposed new ideas, the warped path approach, may succeed. What encourages us to pursue constant fields is that for an observer, for whom the energy-momentum tensor $T^\mu_{\ \ \nu}$ of the constant external electromagnetic field is diagonal, we obtain the metric
\begin{equation}
g_{\mu\nu}(\tau) = \eta_{\mu\alpha}\exp\left(2\frac{\tau_0 e^2}{m^2}T^\alpha_{\ \ \nu}\tau \right)\,,
\end{equation}
which exactly reproduces to the first order in $\tau_0$ the Landau-Lifshitz format of the equations of motion.

Another particularly interesting RR study involves the motion of particles in plane wave fields. The well studied case of electron interaction with a light wave edge is here of particular interest~\cite{Hadad:2010mt}. This case was explored using the LL approach, and critical acceleration effects were demonstrated. A self-consistent warped path-metric formulation of RR could lead to directly verifiable experimental outcomes using present day experimental pulsed laser facilities. 

Another possible extension we would like to pursue is the development of a variational principle for RR force  using the here proposed warped path model. Unlike LAD or LL models where a variational principle was never established, the warped path formulation has a better chance of arising from a specific covariant action since the resultant equations of motion do not contain higher order derivative terms.

{\bf To conclude:}
We have shown that it is possible to describe mechanically decelerated particle energy loss due to radiation friction without introducing the Schott force term and instead we proposed warped matter modification along the particle's path. Our approach resolves well known contradictions: For example, the Landau-Lifshitz-like procedure predicts that particles would gain energy, presumably due to interaction with its own radiation field~\cite{Bild:2019dlu}. The  warped path approach does not introduce such an interaction and the total energy loss remains consistent with the Larmor radiation energy loss formula. Therefore a conceptual advantage of our proposed reformulation of RR is that we do not need to provide an interpretation of the causality difficulties created by the contorted derivation and implementation of LAD.  There is no self-acceleration without external force possible in our approach.

We have shown that when solved consistently, radiative fields are given by particle acceleration due to both external force and radiation friction. The prediction of such a self-consistent calculation is that the radiation friction at most doubles the \lq mechanical\rq\ energy loss, see Figure \ref{fig:delta}. This result is intuitively and theoretically satisfactory and it can have some interesting experimental consequences awaiting study in parton jet quenching processes in quark-gluon plasma.

\acknowledgements{Martin Formanek and Johann Rafelski would like to express their gratitude to Dr. Tam\'as Bir\'o and Dr. P\'eter L\'evai for their hospitality during the summer 2019 at the Wigner Research Centre for Physics, Budapest and 2019 Balaton Workshop when part of this work was conducted. Johann Rafelski was a Fulbright Fellow during this period.

\end{document}